\documentclass[10pt,a4paper]{article}

\usepackage[a4paper, margin=1in]{geometry}
\usepackage{authblk}
\usepackage{amssymb}
\usepackage{hyperref}
\usepackage{graphicx}
\usepackage{epsfig}
\usepackage{epstopdf}
\usepackage{array}
\usepackage{siunitx}
\usepackage[usenames, dvipsnames]{color}
\usepackage{color}
\usepackage{subfig}
\usepackage{longtable}
\usepackage[utf8]{luainputenc}

\usepackage{epsfig}
\usepackage{epstopdf}
\usepackage{xspace}
\usepackage{multicol}

\usepackage{makecell}
\usepackage{cite}

\newcommand{\Smilei}{{\sc Smilei}\xspace}
\hypersetup{%
   colorlinks=true,
   urlbordercolor={1 1 1},
   linkcolor={black},
   raiselinks=false,
citecolor=blue,
urlcolor=blue
}

\newcommand\blfootnote[1]{%
  \begingroup
  \renewcommand\thefootnote{}\footnote{#1}%
  \addtocounter{footnote}{-1}%
  \endgroup
}

\begin{document}

\title{\textbf{Efficient cylindrical envelope modeling for laser wakefield acceleration}}

\author{F. Massimo$^1$, I. Zemzemi$^1$, A. Beck$^1$, J. Dérouillat$^2$ and A. Specka$^1$}

\date{\small{$^1$ Laboratoire Leprince-Ringuet – École polytechnique, CNRS-IN2P3, Palaiseau 91128, France\\
$^2$ Maison de la Simulation, CEA, CNRS, Université Paris-Sud, UVSQ, Université Paris-Saclay, 91191 Gif-sur-Yvette, France}}

\maketitle

\blfootnote{*Corresponding author. E-mail address: \href{mailto:massimo@llr.in2p3.fr}{massimo@llr.in2p3.fr}}

\begin{abstract}
The resolution of the system given by Maxwell's equations and Vlasov equation in three dimensions can describe all the phenomena of interest for laser wakefield acceleration, with few exceptions (e.g. ionization). Such arduous task can be numerically completed using Particle in Cell (PIC) codes, where the plasma is sampled by an ensemble of macroparticles and the electromagnetic fields are defined on a computational grid. However, the resulting three dimensional PIC simulations require substantial resources and often yield a larger amount of information than the one necessary to study a particular aspect of a phenomenon. Reduced models, i.e. models of the Maxwell-Vlasov system taking into account approximations and symmetries, are thus of fundamental importance for preliminary studies and parametric scans. In this work, the implementation of one of these models in the code \Smilei, an envelope description of the laser-plasma interaction with cylindrical symmetry, is described.
\end{abstract}

\section{Introduction}
Laser Wakefield Acceleration (LWFA) \cite{TajimaDawson79,Malka2002,Esarey2009,Malka2012} consists in the excitation of a plasma electron wave in the wake of an intense laser pulse propagating in an underdense plasma. The longitudinal electric field in this plasma wave can be of order of magnitudes higher than the accelerating fields in metallic cavities of conventional accelerators. Electrons injected with sufficient velocity and in the proper phase of the plasma wave can be accelerated to relativistic energies as they propagate with the laser creating the wave. Typical carrier wavelengths $\lambda_0$ of lasers used to realize this kind of particle acceleration are of the order of one micron, while the plasma wavelength in typical regimes of LWFA is of the order of tens or hundreds of microns. The laser pulse envelope is normally resonant with the plasma wavelength, so it has often a length and waist size of the order of the plasma wavelength. The numerical method of choice to simulate LWFA is the Particle in Cell (PIC) method \cite{BirdsallLangdon2004}, where the plasma is discretized through an ensemble of macroparticles and the electromagnetic fields are defined on a grid that discretizes the physical space. The Maxwell-Vlasov system of equation is then solved self-consistently. Given the disparity between the largest length to simulate with the PIC method (the plasma wavelength and the accelerator length, normally longer than one millimeter) and the smallest scale to resolve (the laser wavelength), PIC simulations of LWFA in three dimensions (normally required for the desired physical accuracy \cite{Davoine2008}) need a considerable amount of resources. Reduced models exploiting physical approximations or symmetries are thus of paramount importance for preliminary studies and parametric scans necessary to model a LWFA experiment. One of these approximation is the envelope or ponderomotive approximation \cite{Mora1997,Gordon2000,Huang2006,Cowan2011,Benedetti2012,Helm2016}, where the laser is described through the complex envelope of its vector potential, eliminating the need to resolve its high frequency oscillations. Its large space-time scale interaction with the plasma is then described through the ponderomotive force of the laser acting on the plasma particles, which for their part influence the laser propagation with their susceptibility. Recently, an implementation of a simple, easily parallelizable envelope model has been proposed \cite{Terzani2019} and applied by the authors \cite{Massimo2019} to set ups of interests for Apollon \cite{Cros2014} in its 3D formulation. Despite its inherent considerable reduction of the necessary computation time, the envelope model speed can be further increased taking advantage of the cylindrical symmetry of many physical set ups of interest. An increasing number of experiments is modeled through cylindrical ponderomotive PIC codes, like INF\&RNO \cite{Benedetti2012} and Osiris \cite{Helm2016}, with applications for example to the LWFA experiments in BELLA \cite{Gonsalves2019} and the laser-induced ionization in the AWAKE experiment \cite{AWAKE} respectively. The simulation of these experiments would be significantly more costly (or even impossible in the case of AWAKE) without the envelope approximation or the use of cylindrical symmetry. In this paper, the implementation of the envelope model presented in \cite{Terzani2019} with cylindrical symmetry in the PIC code \Smilei \cite{Smilei2018,Beck2019} is presented. In the second section, the model equations are reviewed. In the third section, their numerical solution is described and in the fourth section three benchmarks of this implementation are shown.

\section{Review of the envelope model in cylindrical symmetry}
In the following equations, normalized units will be used. The electric charge is normalized by the elementary charge, the speed by the speed of light, the mass by the electron mass and frequencies by the laser frequency $\omega_0 = 2\pi c/\lambda_0$.
As described in \cite{Mora1997,Cowan2011}, the fundamental assumption of an envelope model is to describe the laser vector potential as a slowly varying complex function $\tilde{A}$ modulated by a carrier frequency $\omega_0$, propagating in the $x$ direction:
\begin{equation}\label{envelope_form}
\hat{A}(\mathbf{x},t)=\textrm{Re}\left[\tilde{A}(\mathbf{x},t)e^{i(x-t)}\right],
\end{equation}
D'Alembert's Equation applied to $\hat{A}$ can then be rewritten as function of the laser envelope $\tilde{A}$ and using the assumption of cylindrical symmetry, i.e. the azimuthal derivatives $\partial_\theta$ equal to zero:
\begin{equation}\label{envelope_equation}
\partial^2_x \tilde{A}+ \frac{1}{r}\partial_r\left(r\partial_r \tilde{A} \right)+2i\left(\partial_x \tilde{A} + \partial_t \tilde{A}\right)-\partial^2_t\tilde{A}=\chi \tilde{A}.
\end{equation}
The susceptibility $\chi$ quantifies the response of the plasma to the laser, modifying the propagation of the laser itself \cite{Cowan2011}.
The evolution of the ``low frequency" electromagnetic fields (denoted with a bar) can be described through Maxwell's equations written using the assumption of cylindrical symmetry (see \cite{MassimoJCP2016} or \cite{Lifschitz2009}, from the latter taking only the azimuthal mode 0 , which corresponds to perfect cylindrical symmetry):
\begin{equation}\label{MaxwellEqsCylindrical}
\partial_t  \bar{B}_{\theta}= \partial_{r}\bar{E}_{x} -\partial_{x}\bar{E}_{\rm r}, \quad \partial_t  \bar{E}_{ r} = -\partial_x  \bar{B}_{\theta}-\bar{J}_r, \quad \partial_t \bar{E}_{x}=  \frac{1}{r}\partial_r\left(r\bar{B}_{\theta}\right)-\bar{J}_x.\\
\end{equation} 
As described in \cite{Mora1997,Cowan2011} under the envelope assumption given by Eq. \ref{envelope_form}, the equations of motion of the particles are then modified to include a ponderomotive term, i.e. the averaged effect of the laser oscillations. This term can be expressed only interms of the laser envelope defining the ponderomotive potential $\Phi=|\tilde{A}|^2/2$. The electromagnetic fields in the equations of motion are substituted by their low-frequency counterpart:
\begin{equation}\label{EquationsOfMotion}
\frac{d\mathbf{\bar{x}}}{dt}=\frac{\bar{\mathbf{p}}}{\bar{\gamma}}, \quad \frac{d\mathbf{\bar{p}}}{dt}=-\left(\mathbf{\bar{E}}+\frac{\bar{\mathbf{p}}}{\bar{\gamma}}\times\mathbf{\bar{B}}\right)-\frac{1}{\bar{\gamma}}\nabla\Phi, \quad \bar{\gamma}=\sqrt{1+|\mathbf{\bar{p}}|^2+\Phi},
\end{equation}
where $\mathbf{\bar{x}}$ and $\mathbf{\bar{p}}$ are the averaged positions and momenta of the particles. From the definition of the ponderomotive Lorentz factor $\bar{\gamma}$, the susceptibility $\chi$ in Eq. \ref{envelope_equation} in a given point $x$ is defined as the spatial average, defined over the particles in $x$, of their charge density divided by $\bar{\gamma}$.

\section{Numerical implementation}
\begin{figure}
\begin{center}
\includegraphics[width=35pc]{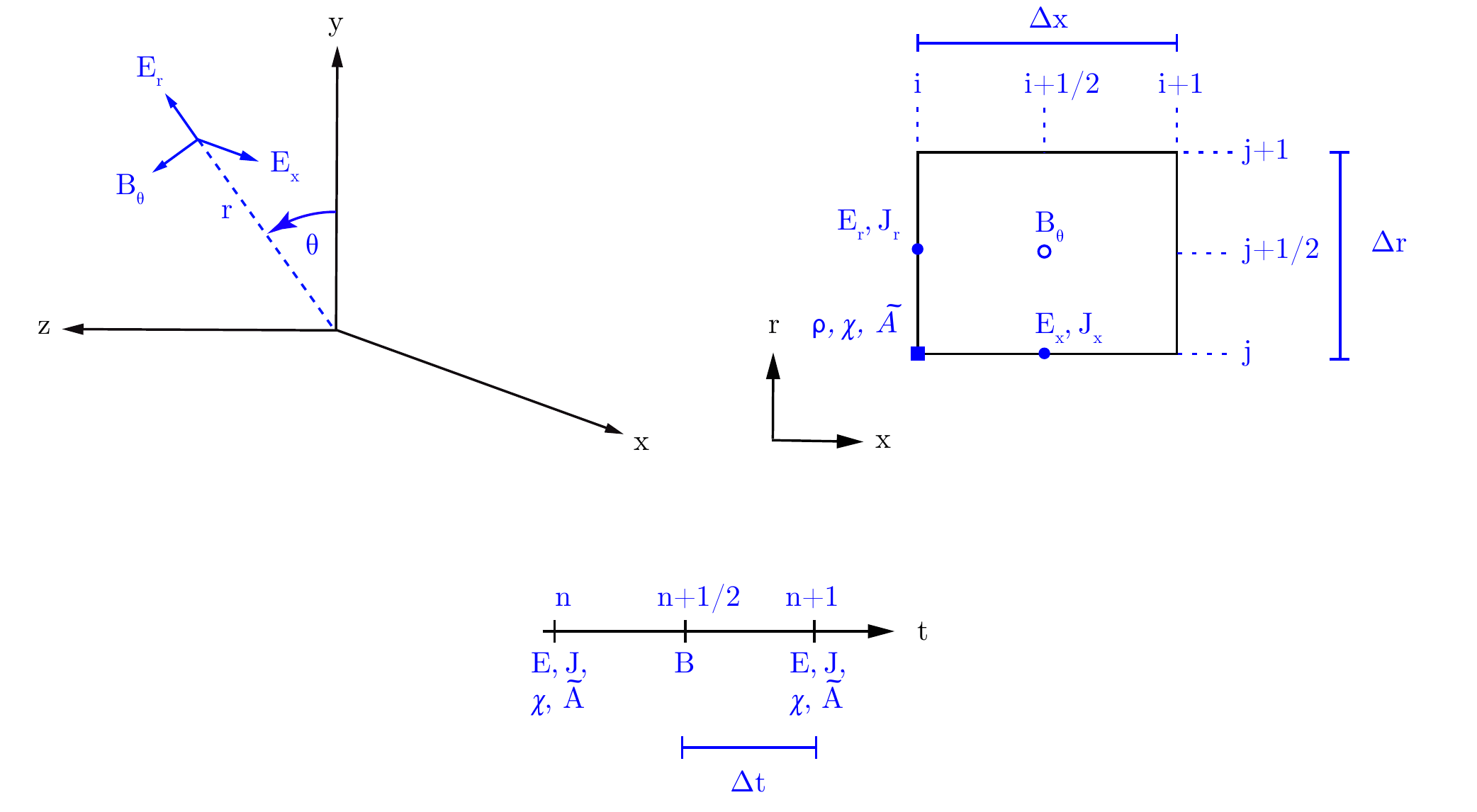}
\end{center}
\caption{\label{mesh_coordinate_reference} Top left: reference axes and definition of the cylindrical electromagnetic fields. Top right: spatial centering of the grid quantities. Bottom: time centering of the grid quantities.}
\end{figure}

The sequence of operations needed to solve the equations in cylindrical symmetry outlined in the previous section is the same as the one used to solve the model equations in 3D, explained in \cite{Terzani2019,Massimo2019}. The major differences rely in the spatial discretiziation of the physical quantities on the grid in cylindrical geometry and the correspondent solvers. Although the fields are defined on a 2D grid, the particles positions and momenta are defined in the 3D space. 
The coordinate axes used in the cylindrical geometry of \Smilei are defined as in Fig. \ref{mesh_coordinate_reference}. The electromagnetic fields are defined in the cell borders and interior according to a Yee-like cell for the Finite Difference Time Domain solver for Maxwell's equations \cite{Yee1966}. The spatial centering of the electromagnetic fields follows \cite{Lifschitz2009}, shown in Fig. \ref{mesh_coordinate_reference} with the centering of the envelope quantities (envelope and susceptibility). The envelope quantities are centered in time like the electric field in the Yee scheme (see Fig. \ref{mesh_coordinate_reference}).  
Using centered finite differences to discretize the derivatives in the envelope equation \ref{envelope_equation}, the cylindrical solver is obtained:
\begin{eqnarray}\label{envelope_solver}
\tilde{A}_{ij}^{n+1}= \frac{1+i\Delta t}{1+\Delta t^2} \Bigg[2 \tilde{A}_{ij}^n-(1+i\Delta t)\tilde{A}_{ij}^{n-1}+\Bigg(\nabla^2\tilde{A}|_{ij}^n-\chi^n_{ij}\tilde{A}_{ij}^n+2i\Delta t^2 \frac{\tilde{A}_{i+1\thinspace j}^{n}-\tilde{A}_{i-1\thinspace j}^{n}}{2\Delta z}\Bigg) \Bigg], \quad
\end{eqnarray}
where
\begin{eqnarray}
\nabla^2\tilde{A}\bigg|_{ij}^n=&\frac{\tilde{A}_{i+1 \thinspace j}^{n}-2\tilde{A}_{ij}^{n}+\tilde{A}_{i-1\thinspace jk}^{n}}{\Delta x^2}+\frac{\tilde{A}_{ij-1\thinspace }^{n}-2\tilde{A}_{ij}^{n}+\tilde{A}_{ij+1\thinspace}^{n}}{\Delta r^2}+\frac{1}{r_j}\frac{\tilde{A}_{ij+1}^{n}-\tilde{A}_{ij-1}^{n}}{2\Delta r}.
\end{eqnarray}
The integration timestep, longitudinal and radial mesh cell size are denoted with $\Delta t$, $\Delta x$ and $\Delta r$ respectively. The subscripts containing $i$ and $j$ denote the longitudinal and radial indices on the grid and the subscripts containing $n$ the time indices, as in Fig. \ref{mesh_coordinate_reference}. The susceptibility $\chi^n_{ij}$ is projected from the particles on the grid similarly to the current and charge density \cite{Terzani2019,Massimo2019}.
 
The cylindrical components of the gradient of the ponderomotive potential $\Phi$, i.e. $\partial_x \Phi$ and $\partial_r \Phi$ ($\partial_\theta\Phi=0$  in cylindrical symmetry) are computed through centered finite differences, to be used in the equations of motions of the particles. To move the particles in the 3D space, the 3D cartesian components of the force acting on the particle are computed, then the modified Boris pusher described in \cite{Terzani2019,Massimo2019} is used to solve the particles equations of motion in terms of the 3D cartesian components of positions and momenta. The final form of the equations of motion read, for electrons:
\begin{eqnarray}\label{motion_equations_cartesian}
\mathbf{\bar{x}}=[\bar{x},\bar{y},\bar{z}],\quad \mathbf{\bar{p}}=[\bar{p}_x,\bar{p}_y,\bar{p}_z],\quad cos(\theta) = \bar{y} / \sqrt{\bar{y}^2+\bar{z}^2},\quad sin(\theta) = \bar{z} / \sqrt{\bar{y}^2+\bar{z}^2},\\
\frac{d\mathbf{\bar{x}}}{dt}=\frac{\bar{\mathbf{p}}}{\bar{\gamma}}, \quad\bar{\gamma}=\sqrt{1+|\mathbf{\bar{p}}|^2+\Phi},\nonumber \\
\frac{d\bar{p}_x}{dt}=-\left(\bar{E}_x-\frac{1}{\bar{\gamma}}\partial_x\Phi\right)+\frac{\bar{B}_\theta}{\bar{\gamma}}\left[\bar{p}_ycos(\theta)-\bar{p}_zsin(\theta)\right],\nonumber \\
\frac{d\bar{p}_y}{dt}=-\left(\bar{E}_r+\frac{1}{\bar{\gamma}}\partial_r\Phi + B_{\theta}\frac{\bar{p}_x}{\bar{\gamma}} \right)cos(\theta),\nonumber\\
\frac{d\bar{p}_z}{dt}=\left(\bar{E}_r+\frac{1}{\bar{\gamma}}\partial_r\Phi + B_{\theta}\frac{\bar{p}_x}{\bar{\gamma}} \right)sin(\theta).\nonumber
\end{eqnarray}
The 3D cartesian modified Boris pusher described in \cite{Terzani2019,Massimo2019} is used to solve Eqs. \ref{motion_equations_cartesian}, to avoid the decrease in accuracy observed in the cylindrical version of the Boris pusher \cite{Delzanno2013}.
Maxwell's equations (Eq. \ref{MaxwellEqsCylindrical}) are solved following the FDTD method \cite{Yee1966}, i.e. discretizing the space and time derivatives with centered finite differences, with a $\Delta t/2$ staggering in time. 
The susceptibility is projected as described in \cite{Terzani2019,Massimo2019}, but on the cylindrical grid (see Fig. \ref{mesh_coordinate_reference}). Given the locality of the envelope solver of Eq. \ref{envelope_solver}, the same parallelization strategies used for a standard Yee scheme can be used to exchange the envelope quantities between the sub-domains.
For the simulation of external injection of relativistic particle bunches (see section 3 of \cite{Massimo2019}), the electromagnetic field initialization procedure described in \cite{Vay-PoP2008,Massimo2019} can be easily implemented in cylindrical symmetry, as described in \cite{Massimo2016}.

\section{Benchmarks}
In this section we report three benchmark cases to test the envelope model in cylindrical symmetry. 

\begin{figure}
\begin{center}
\includegraphics[width=38pc]{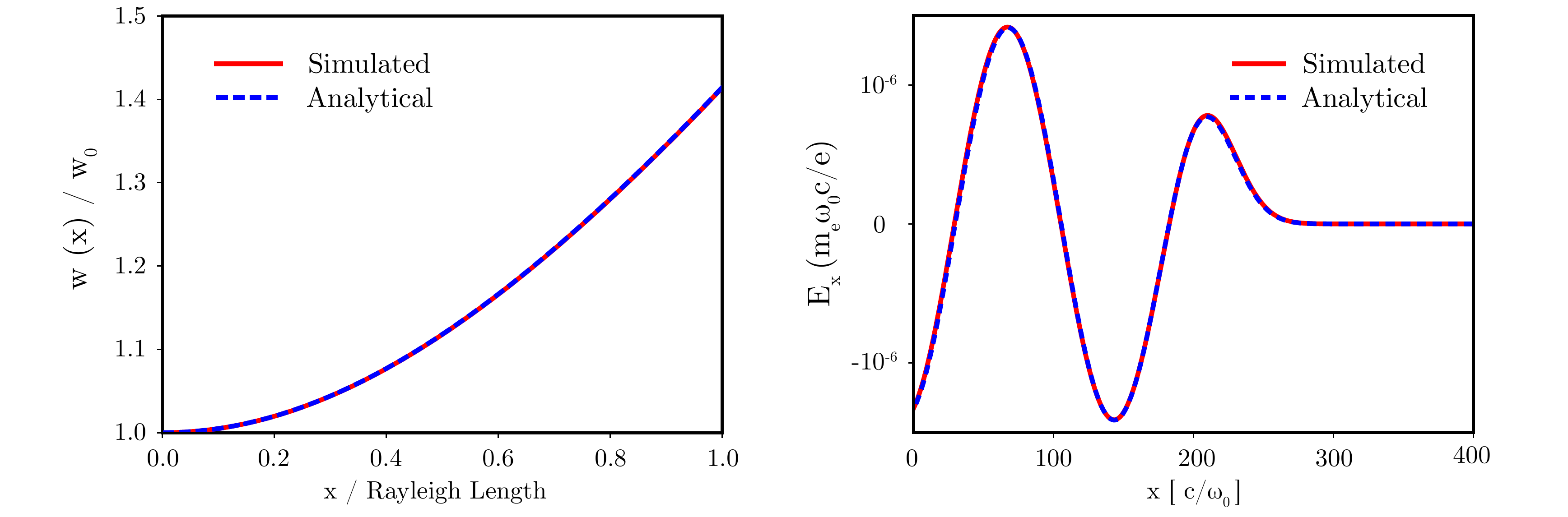}
\end{center}
\caption{\label{vacuum_diffraction_linear_wakefield} Left: Comparison of the simulated and analytical evolution of the waist size $w(x)$ of a Gaussian laser beam. The propagation distance is denoted with $x$. Right: Comparison of the simulated and analytical longitudinal electric field $E_x$ in the wake of a Gaussian laser pulse after a propagation distance $1000$ $c/\omega_0$ in a uniform plasma. The analytical result has been computed from the 1D theory in the linear regime of plasma wake excitation (Eq. 38a of  \cite{Lehe2016FBPIC}).}
\end{figure}

In the left panel of Fig. \ref{vacuum_diffraction_linear_wakefield}, the simulated evolution of the waist size $w(x)$ of a Gaussian laser pulse of initial waist $w_0=90$ $c/\omega_0$ is reported. The laser pulse temporal profile has a FWHM duration $\tau_0=49.5$ $\omega_0^{-1}$ in intensity. The longitudinal and radial resolutions are $\Delta x=0.7$ $c/\omega_0$, $\Delta r =5$ $c/\omega_0$ and the integration timestep is equal to $\Delta t=0.8$ $\Delta x$. The evolution of $w(z)$ follows the analytical Rayleigh's law $w(z)/w_0=\sqrt{1+x^2/(w_0^2/2)}$, where $x$ is the propagation distance. 

The simulation of the excitation of an electron plasma wave in the 1D linear regime is reported as second benchmark. The laser pulse has an initial peak amplitude $a_0=0.01$, initial waist $w_0=90$ $c/\omega_0$ and initial FWHM duration in field $\tau_0=70$ $\omega_0^{-1}$. The plasma has a uniform initial density $n_0= 0.0017$ $n_c$. The longitudinal and radial resolutions are $\Delta x=0.7$ $c/\omega_0$, $\Delta r =5$ $c/\omega_0$ and the integration timestep is equal to $\Delta t=0.8$ $\Delta x$. The plasma is sampled with 6 particles per cell. The computed longitudinal electric field $E_x$ in the wake of a laser pulse compared with its analytical value (Eq. 38a of  \cite{Lehe2016FBPIC}) after a propagation distance $1000$ $c/\omega_0$ in the right panel of Fig. \ref{vacuum_diffraction_linear_wakefield}. A very good agreement with the analytical 1D theory can be inferred. 

\begin{figure}
\begin{center}
\includegraphics[width=38pc]{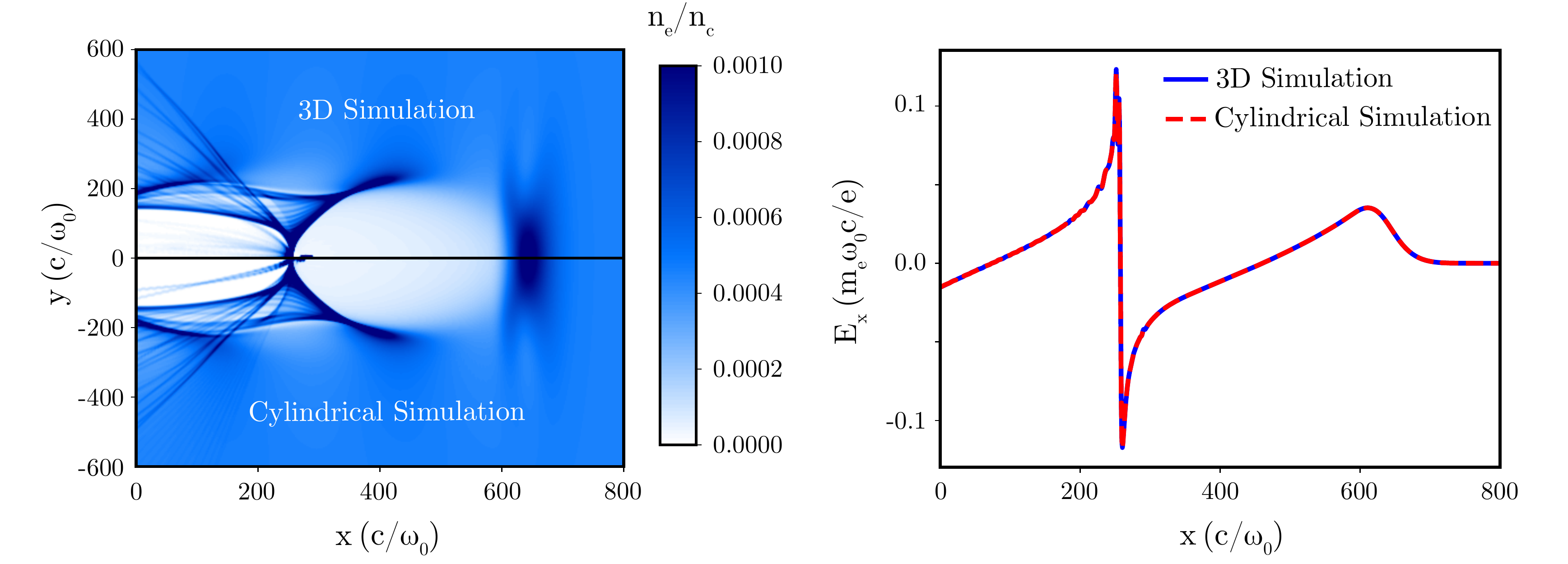}
\end{center}
\caption{\label{Rho2D_and_Ex1D_comparison} Comparison between a 3D envelope simulation and a cylindrical envelope simulation after $7.6$ mm of propagation. Left: snapshot of electron density. Right: longitudinal electric field on the propagation axis. }
\end{figure}

Finally, the preliminary results of a more complex benchmark are reported. The benchmark consists in a nonlinear laser wakefield excitation, with the self-injection and acceleration of an electron beam. The results of a 3D envelope simulation are compared with those of the cylindrical envelope implementation. The laser and simulation parameters of the 3D simulation are those of the ``simulation 4" from \cite{Massimo2019}, reported here for the reader's convenience. A Gaussian laser pulse with waist $40$ $\mu$m, $a_0=2.96$ and FWHM duration in intensity $\tau_0=30$ fs (total energy 15 J) is focused at the start of a plasma with plateau density $n_0=0.8\cdot10^{18}$ cm$^{-3}$. Since the physical setup is cylindrically symmetric, we expect very similar results with a cylindrical implementation of the envelope model. The plasma is sampled with 8 and 6 particles per cell in the 3D and in the cylindrical simulation respectively. In both simulations, the longitudinal mesh cell size and integration timestep are $\Delta x=0.8$ $c/\omega_0$ and $\Delta t = 0.9$ $\Delta x$. The transverse mesh cell size in the 3D simulation and cylindrical simulation are $\Delta y=\Delta z = 3.5 $ $c/\omega_0$ and $\Delta r=3.5$ $c/\omega_0$ respectively. An uncompensated binomial filter (7 passes) \cite{BirdsallLangdon2004,Vay2015} has been applied to the envelope simulation to reduce the effects of the numerical Cherenkov radiation \cite{Lehe2013}. Figure \ref{Rho2D_and_Ex1D_comparison} compares the simulated electron density and the longitudinal electric field on axis after 7.6 mm of propagation. The main features of wake excitation and the injection are well modeled by the cylindrical simulation, with a very good agreement in the longitudinal electric field even after a distance of the order of millimiters. In Table \ref{BeamParams}, the injected electron beam parameters obtained with the two simulations at 7.6 mm (see Table 3 of \cite{Massimo2019} and relative discussion) are reported. A very good agreement is found in the injected beam charge and energy. The beam energy spread is $1.5$ times larger in the cylindrical simulation. This could be caused by the numerical noise on axis, typical of cylindrical PIC simulations, due to high charge macroparticles crossing the propagation axis \cite{Benedetti2010} or to the boundary conditions on the axis. We note a significantly higher emittance in the cylindrical simulation too, probably caused by the same effects. Further studies are necessary to find other possible causes, for example a higher growth rate of numerical Cherenkov radiation \cite{Lehe2013} with a cylindrical FDTD solver.

Normally most of the LWFA  set ups have a near-cylindrical symmetry, i.e. only 2 azimuthal modes are necessary to simulate them \cite{Lifschitz2009}, where the laser is modeled through the azimuthal mode 1. In this particular set up the high frequency oscillations of the laser can be neglected and an envelope model can be used. In this case, the azimuthal mode 1 is not necessary and only the mode 0 (representing the perfect cylindrical symmetry) can be used to perform preliminary scans or data analysis. Thus, a cylindrical envelope model can be of great benefit for the quick study of LWFA, as demonstrated in \cite{Benedetti2010,Benedetti2018}.  In our particular benchmark the cylindrical simulation needed a total of $300$ cpu-hours (distributed on 320 cpus), while the 3D envelope simulation needed a total amount of resources 1200 times greater (both referred to 7.6 mm of propagation). In our case, since the integration timestep, the mesh resolution, the longitudinal and transverse physical dimension of the simulation domain are the same (half transverse domain in the case of the cylindrical simulation), with the hypothesis that the cost scales linearly with the number of particles, the general speedup $S$ can be estimated as 
\begin{equation}\label{Speedup}
S= \frac{N^2_{cells_\perp\thinspace 3D}\times N_{cells_{||}\thinspace 3D}\times N_{ppc\thinspace3D}}{N_{cells_\perp\thinspace cyl}/2\times N_{cells_{||}\thinspace cyl}\times N_{ppc\thinspace cyl}}
\end{equation}
In our case the longitudinal and transverse number of cells of the moving window occupied by the plasma particles is the same, i.e. $N_{cells_{||}\thinspace 3D}= N_{cells_{||}\thinspace cyl} =1088$ and $N_{cells_\perp\thinspace 3D}= N_{cells_\perp\thinspace cyl} =448$ respectively. The number of particles per cell in the two simulations are $N_{ppc\thinspace3D}=8$ and $N_{ppc\thinspace cyl}=6$. With the given values, we indeed obtain an estimated speed-up of $1200$ using Eq. \ref{Speedup}. 

\begin{figure}
\begin{center}
\includegraphics[width=17pc]{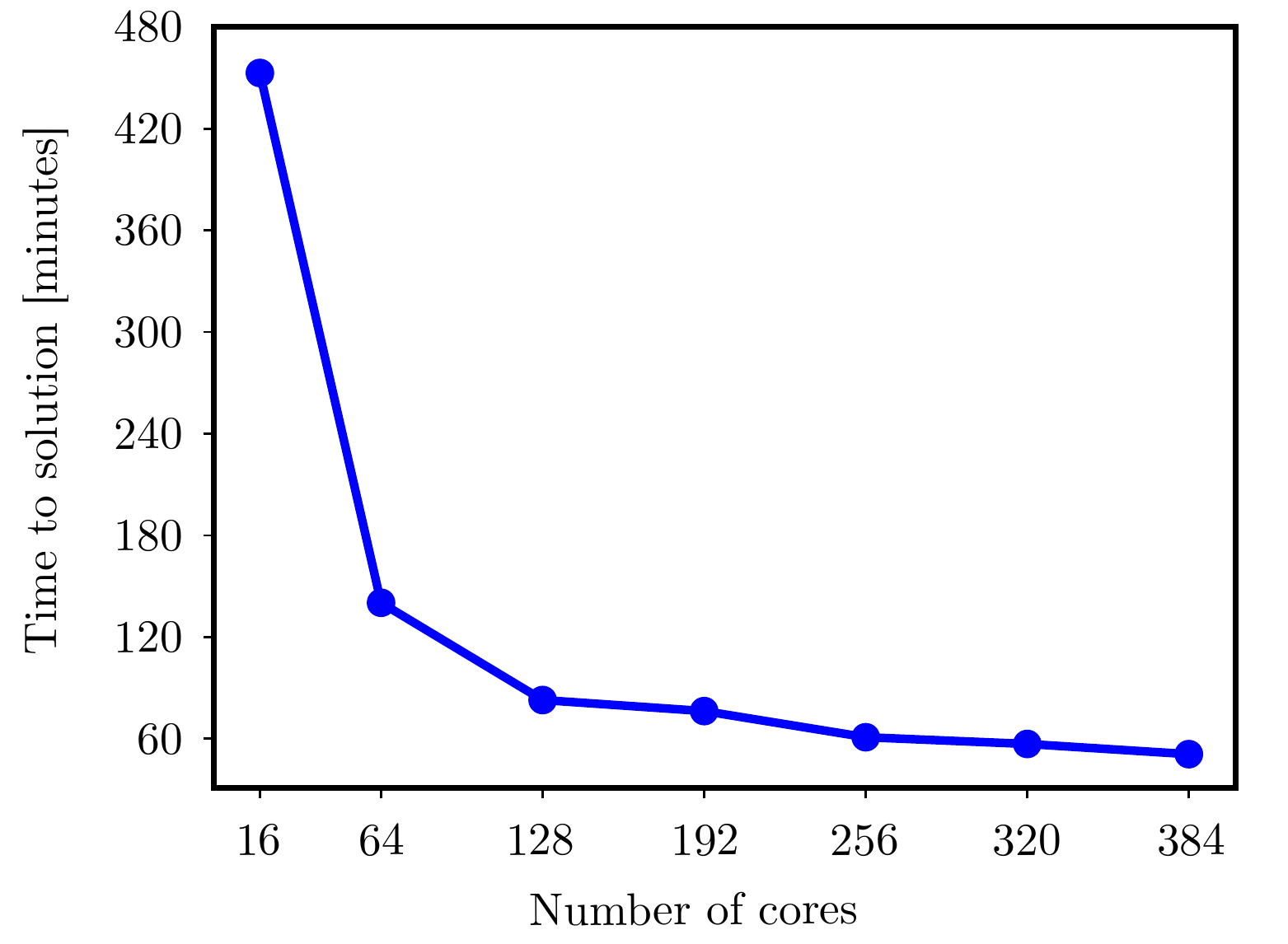}
\end{center}
\caption{\label{time_to_solution} Time needed for 7.6 mm of propagation with thee cylindrical envelope model, varying the number of cores (strong scaling). The tests were made with 16 OpenMP threads for each MPI process. }
\end{figure}

Figure \ref{time_to_solution} reports the time needed by the benchmark simulation (7.6 mm of propagation) varying the number of processors. With 12 standard compute nodes, the time needed is a few minutes per mm of propagation in the plasma.

\begin{center}
\begin{table}[h]
\caption{\label{BeamParams} Injected electron beam parameters after 7.6 mm of propagation: peak energy of the energy spectrum $E_{spectrum\thinspace peak}$, beam charge $Q_{beam}$, relative energy spread $\Delta E/E_{spectrum\thinspace peak}$, normalized transverse emittance $\varepsilon_n=\sqrt{\sigma^2_{x_\perp}\sigma^2_{p_\perp}-\sigma^2_{x_\perp p_\perp}}$, where $\sigma_{x_\perp}$, $\sigma_{p_\perp}$, $\sigma_{x_\perp p_\perp}$ are respectively the rms spread in the transverse position, the rms spread in the transverse momentum and the correlation between the transverse position and the transverse momentum. All these parameters are computed considering the particles within 2 FWHM widths in energy around the spectrum peak)}
\centering
\begin{tabular}{@{}*{7}{l}}
\hline\hline
Simulation&$E_{spectrum\thinspace peak}$[GeV] & $Q_{beam}$[pC]& $\Delta E/E_{spectrum\thinspace peak}$[\%] &$\varepsilon_n$ [mm-mrad]&\\ \hline \hline
3D &1.13&24& 2.0 & 0.4&\\ \hline
cylindrical &1.10&27& 3.2 & 3.5&\\ \hline
\end{tabular}
\end{table}
\end{center}

\section{Conclusions} 
We have described the implementation in cylindrical symmetry in the PIC code \Smilei of the envelope model presented in \cite{Terzani2019}. Its speed allows parametric scans and quick analysis of LWFA experimental set ups where the high frequency oscillations of the laser can be neglected and only their averaged (ponderomotive) effects need to be considered. Preliminary results of comparisons with 3D envelope simulations show an acceptable agreement in the injected electron beam charge and energy, with a total amount of resources smaller by three orders of magnitude.

\section*{Acknowledgements}
F. Massimo was supported by P2IO LabEx (ANR-10-LABX-0038) in the framework “Investissements d’Avenir” (ANR-11-IDEX-0003-01) managed by the Agence Nationale de la Recherche (ANR, France).
This work was granted access to the HPC resources of TGCC under the allocation 2018-A0050510062 made by GENCI. The authors are grateful to the TGCC engineers for their support. The authors thank the engineers of the LLR HPC clusters for resources and help. The authors are grateful to the ALaDyn and \Smilei development teams for the help and discussions during the development of the envelope model, in particular D. Terzani, A. Marocchino, S. Sinigardi, M. Grech, M. Lobet and F. Pérez, and to Gilles Maynard for fruitful discussions.

\bibliography{Bibliography}

\end{document}